\documentclass[review,12pt]{elsarticle}

\usepackage{lineno,hyperref}
\modulolinenumbers[5]

\journal{Journal of \LaTeX\ Templates}

\usepackage{amsmath,amssymb}
\usepackage{graphicx}
\usepackage{epsfig}
\usepackage{color}

\usepackage{caption}
\usepackage{subcaption}

\usepackage{amsfonts}
\usepackage{bm}

\newcommand\be{\begin{equation}}
\newcommand\ee{\end{equation}}
\newcommand\ba{\begin{eqnarray}}
\newcommand\ea{\end{eqnarray}}

\renewcommand{\(}{\left(}
\renewcommand{\)}{\right)}
\renewcommand{\[}{\left[}
\renewcommand{\]}{\right]} 

\newcommand\de{\delta}

\newcommand{\vect}[1]{\bm{\mathrm{{#1}}}}
\newcommand{\fnl}{f_{{\rm NL}}}
\newcommand{\sz}{{(s)}}
\newcommand\phs{\phi_*}
\newcommand\phu{\phi_u}
\newcommand\chs{\chi_*}
\newcommand\chu{\chi_u}
\newcommand\pa{\partial}

\newcommand\e{{\rm e}}
\newcommand\lm{\lambda}
\newcommand\fp{f_{\phi}}

\begin{document}
\begin{frontmatter}
\title{Squeezed bispectrum from multi-field inflation with curved field space metric}

\author{Sakdithut Jitpienka}
\address{The Institute for Fundamental Study \lq\lq The Tah Poe Academia Institute\rq\rq, \\Naresuan University, Phitsanulok 65000, Thailand}
\author{Khamphee Karwan}
\address{The Institute for Fundamental Study \lq\lq The Tah Poe Academia Institute\rq\rq, \\Naresuan University, Phitsanulok 65000, Thailand}
\address{Thailand Center of Excellence in Physics, Ministry of Education,
Bangkok 10400, Thailand}

\begin{abstract}
We investigate influences of the curved field-space metric of multi-field inflationary models on the squeezed bispectrum.
The reduced bispectrum in squeezed limit is computed using the $\delta N$ formalism.
The calculation is performed under the slow-roll approximation and assumption that
field derivative of the field-space metric is sufficiently small such that the contributions from Riemann tensor of the field-space can be approximately ignored.
Based on these approximations, We compute the analytic expressions for the reduced bispectrum in squeezed limit,
and find that, for such a nearly flat field-space metric, the field dependence of the metric can significantly alter both amplitude and shape of the reduced bispectrum.
The reduced bispectrum from this nearly flat field-space metric can lead to spectral index of the halo bias which amplitude is 2 -- 4 times larger than that from the flat field-space model.
This modification of the spectral index of the halo bias due to the curved field-space metric could leave observable imprints in future galaxy surveys.
 \end{abstract}


\end{frontmatter}


\section{Introduction}

Cosmological inflation \cite{Guth:1980zm, Linde:1981mu, Albrecht:1982wi} is a successful paradigm for describing homogeneity and isotropy of the observable universe on large scales as well as the generation of the primordial cosmological perturbations which are seeds of large scale structures in the universe.
Since inflation proceeds at energy scales which can not be recently reached in laboratories,
inflationary models cannot be tested in laboratories and consequently varieties of inflationary models have been proposed from various theories of high energy physics and gravity.
However, physical properties of inflationary models are encoded in predicted primordial perturbations which can be probed in observations of Cosmic Microwave Background \cite{CMB-1, CMB-2, CMB-3} and large scale structure in the universe \cite{LSS-1, LSS-2}.
Hence, features of the primordial perturbations can be used to discriminate between inflationary models and also test physics of the early universe.

In principle, there should exist a huge number of scalar fields in the early universe as predicted by theories of high energy physics such as supersymmetry or string theory \cite{string1, string2}.
The kinetic term of scalar fields arising in high energy theory can involve a non-trivial metric of field-space \cite{Nilles:1983ge}.
Thus it is reasonable to study inflationary models in which inflation is driven by multiple scalar fields with curved field-space metric.
Effects of the field-space metric on power spectrum and its spectral index have been studied in \cite{Sasaki:1995aw, Nakamura:1996da, Gong:2002cx}.
It has been shown that the curved field-space metric leads to a new contribution on spectral index in terms of the field-space Riemann tensor which can be order of usual slow-roll parameter.
Bispectrum from multi-field inflationary model with curved field metric has been investigated in \cite{Gong:2011uw, Elliston:2012uy}
using covariant formulae for field perturbations and extenstion of the $\delta N$ formalism~\cite{Sasaki:1995aw,Nakamura:1996da,Gong:2002cx,Starobinsky:1986fxa}.
Based on the covariant formulae, scale dependence of reduced bispectrum receives contributions from the field-space metric in terms of field-space Riemann tensor and its derivatives \cite{Byrnes:12}.

Due to the conservation of momentum, tree wave vectors of the perturbation modes relevant to the bispectrum  sum to zero, i.e.,
they form a triangle in Fourier space.
For a simplest configuration, these wave vectors have the same length so that the equilateral triangle is formed,
and hence the bispectrum is in the equilateral limit.
Studies of bispectrum from inflationary models with non-trivial field metric mentioned in the previous paragraph have been performed in the equilateral or near equilateral limit.
Here, we are interested in the bispectrum in the squeezed limit at which  length of one wave vector is much smaller than the others.
In our calculation, we use slow-roll approximation and suppose that slope of the  field metric which is the derivative of the field metric with respect to fields is significantly small such that the contributions from the Riemann tensor of the field-space can be negligible.
In order to investigate effects of curved field-space metric on squeezed bispectrum,
it is more convenient to treat terms in the Lagrangian arising from field dependent parts of the field metric as usual coupling terms between fields and derivative of the fields \cite{Bruck:14} rather than using covariant formulae for field perturbations.
Based on this point of view, the squeezed bispectrum for the case of curved field metric can be computed using approaches in \cite{Kenton:15, Tada:16} which allow us to study bispectrum in highly squeezed limit.

This article is organized as follows. In Section \ref{sec2},  we review calculation of reduced bispectrum in squeeced limit using $\Delta$N formalism. 
In section \ref{bispec}, we apply formulae in section \ref{sec2} to compute  squeezed bispectrum and associated spectral index of the halo bias for double inflationary model with nearly flat field metric.
We give our conclusion in section \ref{conclusions}.
In appendices, calculations of some terms appearing in reduced bispectrum and spectral index of the halo bias are presented in detail.

\section{Squeeced bispectrum in $\Delta$N formalism}
\label{sec2}
\subsection{$\Delta$N formalism}
\label{sec21}

In this section, we review essential formulae which are required for  calculations of observable quantities from inflationary models using $\Delta$N formalism.
Based on the $\Delta$N formalism, the curvature perturbations on uniform density hypersurfaces on large scales is equivalent
to the perturbations of number of e-foldings $N \equiv \ln(a(t_u, \vect{x}) / a(t_*))$ needed to foliate from spatially flat hypersurfaces at time $t_*$ to uniform density hypersurfaces at time $t_u$.
Here, $a(t)$ is a cosmic scale factor of the Friedmann-Lema\^{i}tre-Robertson-Walker (FLRW) metric describing global expansion of the background universe,
while $a(t, \vect{x}) \equiv a(t) \e^{\psi(t,\vect{x})}$, where $\psi(t, \vect{x})$ is the perturbation in spatial metric, presents the local expansion of the universe at $\vect{x}$ on large scales.
Using the separate universe approach and slow-roll approximation, a local expansion of the universe can be written in term of the scalar fields $\varphi^I$ as $a(t, \vect{x}) = a(\varphi^I(t, \vect{x}))$ \cite{Lyth:1984gv,Wands2000}.
Thus the curvature perturbation on uniform density hypersurfaces can be expressed in terms of the perturbations in inflaton fields in slow roll limit as \cite{Sasaki:1995aw, Nakamura:1996da, Gong:2002cx}
\be
\zeta(t_u, \vect{x})  = \de N(a(t_u, \vect{x}), a(t_*))
= N_I(t_u, t_*) \delta\varphi^I(t_*, \vect{x}) \, + \dots \,,
\label{zdeltan}
\ee
where Latin indices $I, J, K$ run over the field components,
$\de \varphi^I$ denotes field perturbations on spatially flat hypersurfaces
and $N_I(t_u, t_*) \equiv \partial N(t_u, t_*) / \partial \varphi^I(t_*)$ is  the derivative of the e-foldings of the background universe.
The features of the curvature perturbations predicted from inflationary models can be presented in terms of the spectra which quantify the n-point correlation functions of the curvature perturbations in Fourier space as
\ba
\langle \zeta_{\vect{k_1}}\zeta_{\vect{k_2}} \rangle &=& P_{\zeta}(k_1)  (2 \pi)^3 \delta(\vect{k_1} + \vect{k_2})\,,
\\
\langle \zeta_{\vect{k_1}}\zeta_{\vect{k_2}}\zeta_{\vect{k_3}} \rangle &=&  B_{\zeta}(k_1,k_2,k_3)(2 \pi)^3  \delta(\vect{k_1} + \vect{k_2} + \vect{k_3})\,,
\label{def-bispec}
\ea
where $\langle \dots \rangle$ denotes an ensemble average,
$P_{\zeta}(k_1)$ is the power spectrum and 
$B_{\zeta}(k_1,k_2,k_3)$ is the bispectrum.
From (\ref{zdeltan}), the power spectrum of the curvature perturbations can be computed from the inflaton perturbations on spatially flat hypersurfaces as
\be
P_{\zeta}(k_1) \simeq N_I N_J P^{IJ}(k_1)\,,
\quad
\mbox{where}
\quad
\langle \delta\varphi^I_{\vect{k_1}} \delta\varphi^J_{\vect{k_2}} \rangle = P^{IJ}(k_1)  (2 \pi)^3 \delta(\vect{k_1} + \vect{k_2})\,.
\label{pzpp}
\ee

\subsection{Squeeced bispectrum}
\label{sec22}

From the definition of bispectrum given in Eq.~(\ref{def-bispec}),
the constraint from the delta  function suggests that the wave vectors $\vect{k}_1, \vect{k}_2$ and $\vect{k}_3$ form a closed triangle.
One of a possible configurations of the wave vector is a equilateral  triangle which $k_1 = k_2 = k_3$.
This configuration is usually used in the calculation of the bispectrum by $\Delta$N formalism.
However, in this work, we are interested in the bispectrum in squeezed limit at which $k_3 \approx k_2 \gg k_1$, so that the wave vector forms a ``squeezed'' triangle.

In order to quantify magnitude of non-Gaussianity,
the reduced bispectrum $\fnl$ is defined from the ratio between the bispectrum and the square of the power spectrum as
\be
\frac 35 \fnl(k_1,k_2,k_3) \equiv \frac 12 \frac{B_{\zeta}(k_1,k_2,k_3)}{[P_{\zeta}(k_1)P_{\zeta}(k_2)+ (k_1 \to k_2 \to k_3  )]}\,.
\label{fnl}
\ee
In the squeezed limit, the above equation becomes
\be
\lim _{k_1 \ll k_2 \sim k_3} \frac{3}{5}\fnl(k_1,k_2,k_3) \equiv 
\frac 35 \fnl^\sz (k_1,k_2)
\simeq  \frac{\lim _{k_1 \ll k_2} B_{\zeta}(k_1,k_2,k_2)}{4 P_{\zeta}(k_1) P_{\zeta}(k_2)}\,.
\label{fnl-sq}
\ee
The main contribution to the bispectrum in the squeezed limit comes from the correlation between the long wavelength perturbations and
the power spectrum of the short wavelength perturbations on large scales \cite{Kenton:15, Tada:16}.
This correlation is a consequence of the modulation of the amplitude of the short wavelength perturbations
by the long wavelength perturbations when the short wavelength perturbations exit the Hubble radius.
Based on this conclusion, the squeeced bispectrum can be computed in $\Delta$N formalism using the relation \cite{Tada:16,Kenton:16}
\be
\frac{3}{5}\fnl^\sz(k_1, k_2)
= \frac{N_I|_L P^{IK}(k_1)|_L}
{4 P_\zeta(k_1)|_L P_\zeta(k_2)|_S}
\left.\frac{\partial P_\zeta (k_2)}{\partial \varphi^J}\right|_S
\left. \frac{\partial \varphi^J_S}{\partial \varphi^K}\right|_L \,.
\label{fnl-sq-ni}
\ee
where subscripts ${}_L$ and ${}_S$ denote evaluation at the time when long wavelength perturbations with wavenumber $k_1$ and short wavelength perturbations with wavenumber $k_2$ exist the Hubble radius respectively.
The perturbations are on spatially flat hybersurfaces when they exit the Hubble radius.
A specification of the Hubble radius exit time depends on definition of the  scales of the perturbations.
In the $\Delta$N formalism,
it is convenient to parameterize time during inflation by the number of e-folding and defined the scale of particular perturbation mode by the number of e-folding at its Hubble radius exit time.
For the forward formulation, 
the scales of two perturbation modes which subsequently exit the Hubble radius are defined by the number of e-folding realised between their Hubble radius exit \cite{Tada:16}.
Alternatively, in the backward formulation, the scale of particular perturbation mode are defined by the number of e-folding
realised backwards in time between the end of inflation and a time at which the perturbation exits the Hubble radius.
In the following calculations, the number of e-foldings at which the long and short wavelengths perturbation modes exit the Hubble radius are denoted by $N_L$ and $N_S$, respectively.
We suppose that the uniform density hypersurface is reached at the end of inflation at $N = N_u$.

   \section{Squeezed bispectrum in multi-inflaton with curved field-space}
\label{bispec}

We consider the multi-field inflation described by the action
\be
S = \int d^4x \sqrt{-g}\[\frac 12 R
- \frac 12 G_{IJ} \partial_\mu \varphi^I \partial^\mu \varphi^J - W(\varphi^I),
\]\,,
\label{act}
\ee
where the reduced Planck mass is set to unity,
$R$ is the Ricci scalar of the spacetime,
$g$ is the determinant of the spacetime metric,
Greek indices run over the spacetime components,
$G_{IJ}$ is a non-trivial  metric of field-space,
and $W(\varphi^I)$ is the potential of inflatons.

In the following consideration, We concentrate  on two inflatons with the additive separable potential
\be
W\(\phi,\chi\) = U\(\phi\) + V\(\chi\)
= \frac 12 m_\phi^2 \phi^2 + \frac 12 m_\chi^2 \chi^2\,,
\label{pot}
\ee
where $m_\phi$ and $m_\chi$ are the mass of fields $\phi$ and $\chi$ which are constant.
In the above equation, we have set $\varphi^I \equiv (\phi, \chi)$.

We choose to work with the metric
\be
G_{IJ} = \(\begin{array}{cc}
1 & 0 \\
0 & G(\phi)\\
\end{array}\)\,,
\qquad
\mbox{where}
\qquad
G(\phi) \equiv \lm_1 + \lm_2 \phi^p\,.
\label{gij}
\ee
Here, $\lm_1, \lm_2$ and $p$ are the constant parameters.
The parameters $\lm_2$ and $p$ parameterize deviation from the trivial constant metric and quantify
slope of the above metric which becomes nearly flat when $\lm_2$ is significantly smaller than unity.
It can be seen from the action (\ref{act}) that variation of $\lm_1$ is equivalent to variation of the ratio $m_\phi / m_\chi$ if $\lm_2 = 0$.
The advantage of using the above form of $G_{IJ}$ is that the constrained equation of two fields can be integrated analytically.
Furthermore, this metric can represent the metrics used in \cite{Bruck:14, Choi:2007s­u, Mizuno:17} in the limit where the magnitude of parameters in those metrics is small.

\subsection{Background evolution}
\label{back}

\begin{figure}
\begin{center}
\includegraphics[width=0.8\textwidth, height=0.8\textwidth]{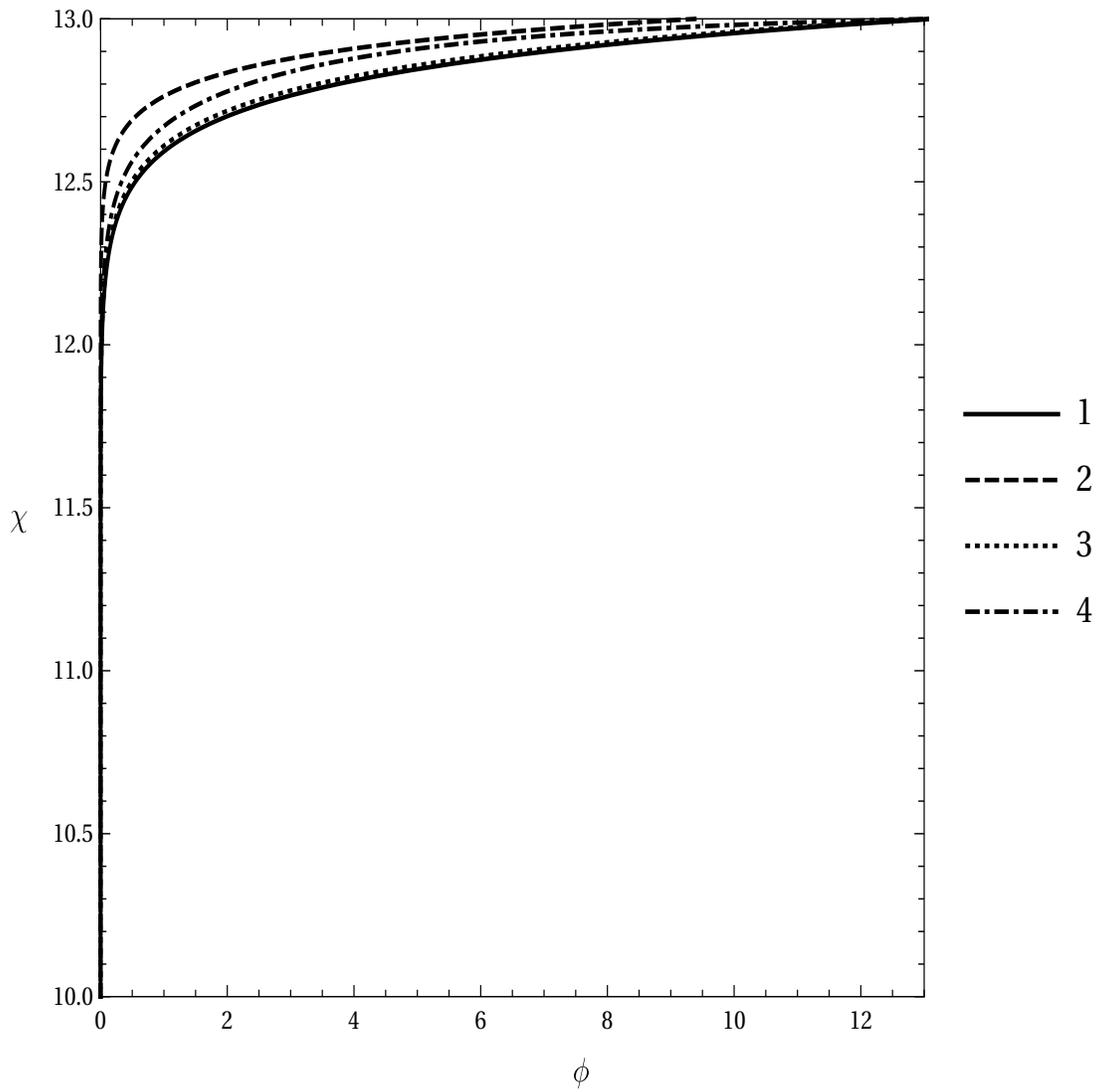}
\caption{\label{fig:1}
Trajectories in the field-space for various values of $\lm_1, \lm_2$ and $p$.
Lines 1, 2, 3 and 4 correspond to the cases where $(\lm_1, \lm_2, p) = (1,0,1), (1.5,0,1), (1,0.01,1)$, and $ (1,0.01,2)$,
respectively.
}
\end{center}
 \end{figure}

Varying the action (\ref{act}) with respect to metric tensor of spacetime,
and inserting the FLRW metric
\be
ds^2 = - dt^2 + a^2 \de_{ij} dx^i dx^j\,,
\label{ds}
\ee
into the result, we obtain Friedmann equation which can be written as \cite{Yokoyama:07}
\be
H^2 \equiv \(\frac{\dot{a}}{a}\)^2
= \frac{2 W}{6 - (\phi')^2 - G(\phi)(\chi')^2}\,,
 \label{h2}
\ee
where a dot denotes derivative with respect to time,
and a prime denotes derivative with respect to number of e-folding of the background universe.
The evolution equation for the background field can be obtained by varying the action (\ref{act}) with respect to the fields,
which yields
\ba
{} && \phi'' - \frac 12 G_\phi (\chi')^2
= -\frac{W}{H^2} \left(\phi' + \frac{U_\phi}{W} \right)\,,
\label{ddphi} \\
{} && \chi'' + \frac{G_\phi}{G(\phi)} \phi' \chi'
= -\frac{W}{H^2} \left( \chi' + \frac{V_\chi}{G(\phi) W} \right)\,,
\label{ddchi}
\ea
where subscripts ${}_\phi$ and ${}_\chi$ denote derivative with respect to $\phi$ and $\chi$ respectively.
Differentiating Eq.~(\ref{h2}) with respect to time,
and inserting $\phi''$ and $\chi''$ from Eqs.~(\ref{ddphi}) and (\ref{ddchi}) into the result,
we can compute slow-roll parameter as
\be
\epsilon \equiv - \frac{\dot H}{H^2}
=
 \frac 12 \((\phi')^2 + G(\phi) (\chi')^2\)\,.
\label{epsi}
\ee
At the lowest order in slow-roll approximation,
Eq.~(\ref{h2}) gives $3 H^2 \simeq W$, and the above equations become
\be
\phi' = -  \frac{U_\phi}W\,,
\qquad
\chi' = - \frac{V_\chi}{G(\phi) W}\,.
\label{kg-slow}
\ee
These equations are valid as long as $|G_\phi| = \lm_2 p \phi^{p -1}$ is sufficiently smaller than unity.
The first integral of these equations gives the following constrained equation:
\be
\int_{\phi_1}^{\phi_2} \frac{d\phi}{G(\phi) U_\phi(\phi)}
= \int_{\chi_1}^{\chi_2} \frac{d\chi}{V_\chi(\chi)}\,,
\label{constr}
\ee
where $(\phi_1, \chi_1)$ and $(\phi_2, \chi_2)$ are any points in field-space.
Substituting the expression for $G(\phi)$ in Eq.~(\ref{gij}) into the above equation,
the relation between the fields $\phi$ and $\chi$ along trajectories which pass a point $(\phi_1, \chi_1)$ in the field-space is
\be
\phi^p = \frac{\lm_1 \fp(\phi_1) \(\chi / \chi_1\)^{r p}}{1 - \lm_2 \fp(\phi_1) \(\chi / \chi_1\)^{r p}}\,,
\label{pofc}
\ee
where $r \equiv \lm_1 m_\phi^2 / m_\chi^2$
and
\be
\fp(\phi_1) \equiv \frac{\phi_1^p}{\lm_1 + \lm_2 \phi_1^p} = \frac{\phi_1^p}{G(\phi_1)}\,.
\label{fect}
\ee
Inserting $\phi'$ and $\chi'$ from Eq.~(\ref{kg-slow}) into Eq.~(\ref{epsi}),
we can write the slow-roll parameter in an approximated form as
\be
\epsilon \simeq
\frac 1{2 W^2} \(U_\phi^2 + \frac{V_\chi^2}{G(\phi)}\)\,.
\label{epsi-slow}
\ee
Using Eq.~(\ref{kg-slow}),
the forward number of e-folding for the background universe realised between times $t_1$ and $t_2$ with $t_2 \geq t_1$ can be computed as
\be
N \equiv \int_{t_1}^{t_2} H d t = - \int_{\phi_1}^{\phi_2} \frac{W(\phi, \chi)}{U_\phi(\phi)} d\phi
= \int_{\phi_2}^{\phi_1} \frac{U(\phi)}{U_\phi(\phi)} d\phi
+ \int_{\chi_2}^{\chi_1} G(\phi(\chi)) \frac{V(\chi)}{V_\chi(\chi)} d\chi\,,
\label{def-n}
\ee
where $(\phi_1, \chi_1)$ and $(\phi_2, \chi_2)$ are points in the field-space at time $t_1$ and $t_2$, respectively.
The backward number of e-folding between times $t_1$ and $t_2$ can  be computed from the above equation by inserting minus sign on the right-hand side of the equation.
Setting $\phi_1$ and $\chi_1$ to be $ \phi_i$ and $\chi_i$ where $\phi_i$ and $\chi_i$ are the initial values of $\phi$ and $\chi$,
the above integration can be expressed in terms of the hypergeometric function ${}_2F_1\(a,b;c;z\)$ as
\begin{align}
N&(\phi,\chi)= \nonumber\\
& \frac 14 \[ \phi_i^2 - \phi^2 
+ \lm_1 \chi_i^2 {}_2F_1\(1, \frac{2}{p r}; 1 + \frac{2}{p r}; \lm_2 \fp(\phi_i)\) 
- \lm_1 \chi^2 {}_2F_1\( 1 , \frac{2}{p r} ; 1 + \frac{2}{p r} ; \( \frac{\chi}{\chi_i} \)^{p r} \lm_2 \fp(\phi_i) \) 
 \]\,.
\label{n-hyper}
\end{align}
In our consideration, the total number of e-folding is always set to 85 and $m_\phi = 9 \times 10^6$GeV $= 9 m_\chi$.
The field $\phi$ dominates dynamics of the universe during initial stage of inflation,
and $\chi$ becomes dominant afterwards until the end of inflation.
For the case of flat field-space, the inflation ends at $\chi = \sqrt{2}$ and $\phi \sim 0$.
In this case, we have $\lm_2 = 0$, so that ${}_2F_1 \simeq 1$, and hence Eq.~(\ref{n-hyper}) yields $\phi_i = \chi_i = 13$.
For the case of the curved field-space, $\chi_i$ is still set to be $13$, while $\phi_i$ is set such that total number of e-folding  is 85.
Since for the case of curved field-space, $\phi$ also drops to zero when $\chi$ becomes dominant,
$G(\phi) \to \lm_1$ and consequently Eq.~(\ref{epsi-slow}) yields $\chi = \sqrt{2 / \lm_1}$ at the end of inflation.
In figure (\ref{fig:1}), we use Eq.~(\ref{pofc}) to plot the trajectories of $\phi$ and $\chi$ in field-space for various values of parameters.
From the plot, we see that $\phi \to 0$ at the end of inflation,
which follows from Eq.~(\ref{pofc}) that if $r \gg 1$, the value of the field $\phi$ at the end of inflation can be extremely smaller than the value at the initial stage of inflation.
From Eq.~(\ref{def-n}), we see that $\lm_1$ can enhance the number of e-folding for a given initial value of $\chi$,
so that the initial value of $\phi$ reduces when $\lm_1$ increases.
The enhancement of $\lm_2$ and $p$ do not significantly alter the initial value of $\phi$.
Since $\phi$ drops towards zero when $\chi$ starts to dominate dynamics of inflation,
the parameters $\lm_2$ and $p$ have no significant effect on dynamics of inflation when $\chi$ completely dominates.
\subsection{Squeezed bispectrum}
\label{calbis}

For multi-field inflationary models, the power spectrum of the field perturbations at the Hubble radius exit in the slow roll approximation is \cite{Sasaki:1995aw}
\be
P^{IJ} \simeq \frac{H^2}2 k^3{} G^{IJ}\,,
\label{p-field}
\ee
where $H$ and $G^{IJ}$ are evaluated at the Hubble radius exit.
Inserting the above equation into Eq.~(\ref{pzpp}) and substituting the result into Eq.~(\ref{fnl-sq-ni}),
we get
\be
\frac{3}{5}\fnl^\sz(N_L, N_S)
= \frac{N_I|_L G^{IJ}(\phi_L)}
{4 (N_I N^I)|_L}
\(\frac{2 N_{K' K} N^{K'} + N_{J'} N_{K'} G^{J' K'}_{,K}}{N_I N^I} + \frac{W_{,K}}{W}\)_S
\left. \frac{\partial \varphi^K_S}{\partial \varphi^J}\right|_L \,.
\label{fnl-sq-ni-rep}
\ee
where a subscript ${}_{,K}$ denotes derivative with respect to $\varphi^K$.
In order to compute $\fnl^\sz$ for model of interests,
we insert the expressions for the potential and metric given in Eqs.~(\ref{pot}) and (\ref{gij}) into the above equation.
We compute  the expressions for $N_I, N_{IJ}$ and $\partial \varphi^J_S/\partial \varphi^K |_L$ in appendix (\ref{appena}).
Since we are interested in $\fnl$ at the end of inflation at which $\phi(t = t_u) \to 0$,
we compute $\fnl^\sz$ for the forward formulation by substituting  Eqs.~(\ref{nisim}), (\ref{nijsim}) and (\ref{gfrep}) into Eq.~(\ref{fnl-sq-ni-rep}), while Eq.~(\ref{gbrep}) is used instead of Eq.~(\ref{gfrep}) for the backward $\fnl^\sz$.
The expressions for $\fnl^\sz$ in the backward  formulation is given by
\be
\fnl{}_b^\sz = 
\frac{D_{S_2} + D_{S_1}\delta_{N_L}}{D_{S_3}}\sum_{i= 1}^4 \(C_{S_{2i-1}} + C_{S_{2i}} G_{\phi_S}\) \delta_{N_S}^{i - 1}\,,
\label{fnlbak}
\ee
where $G_{\phi_S} \equiv \left. \partial G / \partial \phi \right |_S $,
and the expressions for the coefficients $C_{S_i}$ and $D_{S_i}$ are given in appendix (\ref{coeffb}).
The quantity $\delta_{N_S}$ is defined in Eq.~(\ref{delnb}),
while the quantity $\delta_{N_L}$ is also computed from Eq.~(\ref{delnb}) by replacing evaluation at $N_S$ with evaluation at $N_L$.
The expression for $\fnl^\sz$ in the forward  formulation is rather complicated.
However, in our consideration $\fnl^\sz$ in the forward  formulation can still be written in terms of that in the bakward  formulation through the consistency relation,
\be
\frac 35 \fnl{}_f^\sz = \frac 35 \fnl{}_b^\sz + \frac 14 \(1 - n_s\)\,,
\label{consis}
\ee
where $n_s$ is the spectral index of the powerspectrum evaluated at $N_S$,
which is computed as
\be
n_s-1 \equiv \frac{\partial\ln(k^3 P_\zeta)}{\partial N}
= \frac{1}{k^3 P_\zeta} \frac{\partial \varphi^I}{\partial N}\frac{\partial}{\partial \varphi^I}\(k^3 P_\zeta\)
=
- \(\frac{2 N_{K' K} N^{K'} + N_{J'} N_{K'} G^{J' K'}_{,K}}{N_I N^I} + \frac{W_{,K}}{W}\)_S
\left. \frac{W^{,K}}{W} \right|_S\,.
\label{pindex}
\ee
The relation on the third equality agrees with the result in \cite{Sasaki:1995aw, Byrnes:12} at the lowest order in slow-roll parameter when 
$|R_{IJKL}| \ll 1$ for all components.
Here, $R_{IJKL}$ is the field-space Riemann tensor.
In this situation, $\partial G / \partial \phi$ can still have contributions on $n_s$ and $\fnl^\sz$
because $\partial G / \partial\phi$ is larger than $|R_{IJKL}|$ according to the following consideration.
Let us consider the ratio
\be
C \equiv \frac{|R_{IJKL}|}{ \partial G / \partial\phi}\,.
\label{ratioc}
\ee
Based on our field-space metric,
we have $C \sim 1/\phi$ when $p \geq 2$ and $C \sim \lm_2$ when $p =1$ for all components of $R_{IJKL}$.
To ensure that the field-space is nearly flat,
$\lm_2$ is set to be less than unity.
Thus the non-trivial part of the field-space metric can give significant contribution to the observable quantities when $\phi$ is sufficiently larger than unity.
As a result, the ratio $C$ is always less than unity in our consideration.

Inserting the expressions for $N_{IJ}$ and $N_I$ from the appendix (\ref{appena}) into the above equation,
we get
\be
1-n_s = 4\frac{1}{D_n} \sum_{i=1}^3 \(C_{n_{2i -1}} + C_{n_{2i}} G_{\phi_S}\) \delta_{N_S}^{i - 1}\,,
\label{pnsrep}
\ee
where the expressions for the coefficients $D_n$ and $C_{n_i}$ are given in appendix (\ref{coefns}).
To verify Eq.~(\ref{consis}),
we compute $\fnl{}_f^\sz$ by substituting Eqs.~(\ref{fnlbak}) and (\ref{pnsrep}) into Eq.~(\ref{consis}).
We find that $\fnl{}_f^\sz$ obtained from Eq.~(\ref{consis}) is exactly the same as that is directly computed from Eq.~(\ref{fnl-sq-ni-rep}) using Eqs.~(\ref{nisim}), (\ref{nijsim}) and (\ref{gfrep}).

\begin{figure}[!h]
\centering
	\begin{subfigure}[t]{0.45\textwidth}
    \includegraphics[width=\textwidth]{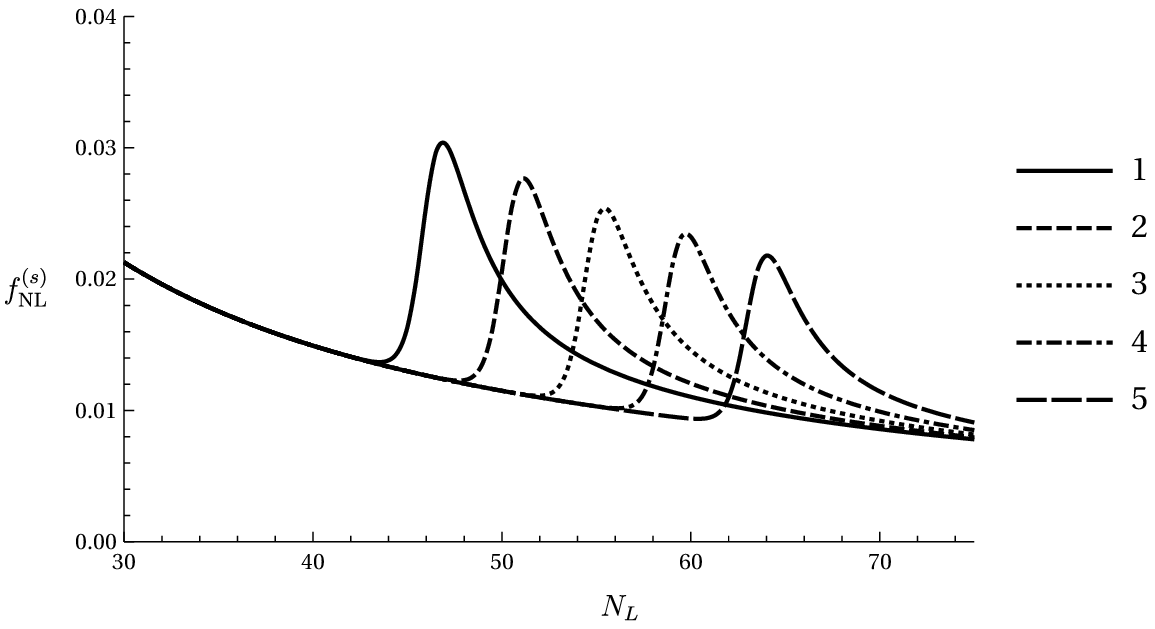}
    \caption{Reduced bispectrum for forward formulation} 
    \label{fnlfvl1}
    \end{subfigure}
    \begin{subfigure}[t]{0.45\textwidth}
    \includegraphics[width=\textwidth]{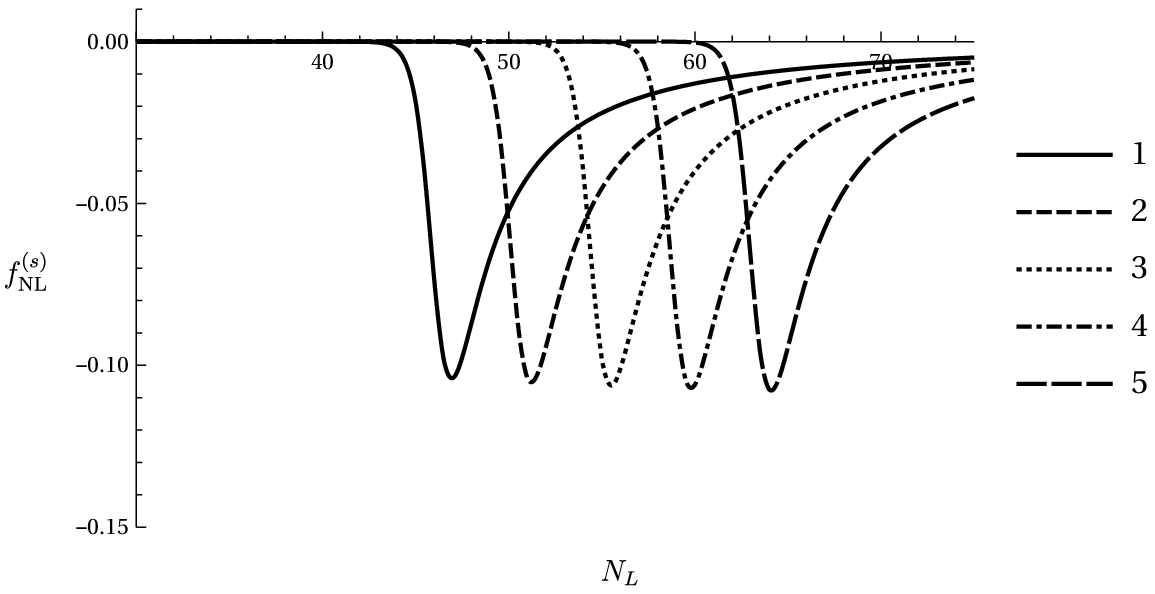}
    \caption{Reduced bispectrum for backward formulation} 
    \label{fnlbvl1}
    \end{subfigure} 
	\caption{
In these plots, lines 1, 2, 3, 4, 5, and 6 correspond to the cases where $ \lm_1$ = 1, 1.1, 1.2, 1.3, 1.4, and 1.5, respectively.
For all lines, $ \lm_2 =0, p=1 $.
} 
\label{fig:varyl1}
\end{figure}

\begin{figure}[!h]
\centering
	\begin{subfigure}[t]{0.45\textwidth}
    \includegraphics[width=\textwidth]{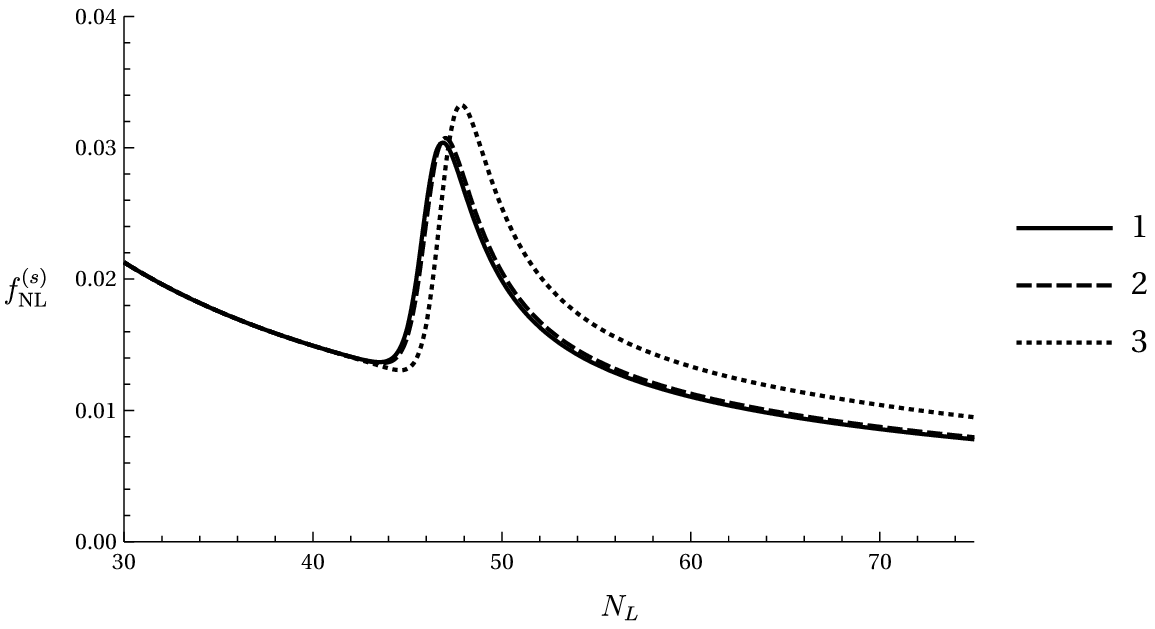}
    \caption{Reduced bispectrum for forward formulation} 
    \label{fnlfvl2}
    \end{subfigure}
    \begin{subfigure}[t]{0.45\textwidth}
    \includegraphics[width=\textwidth]{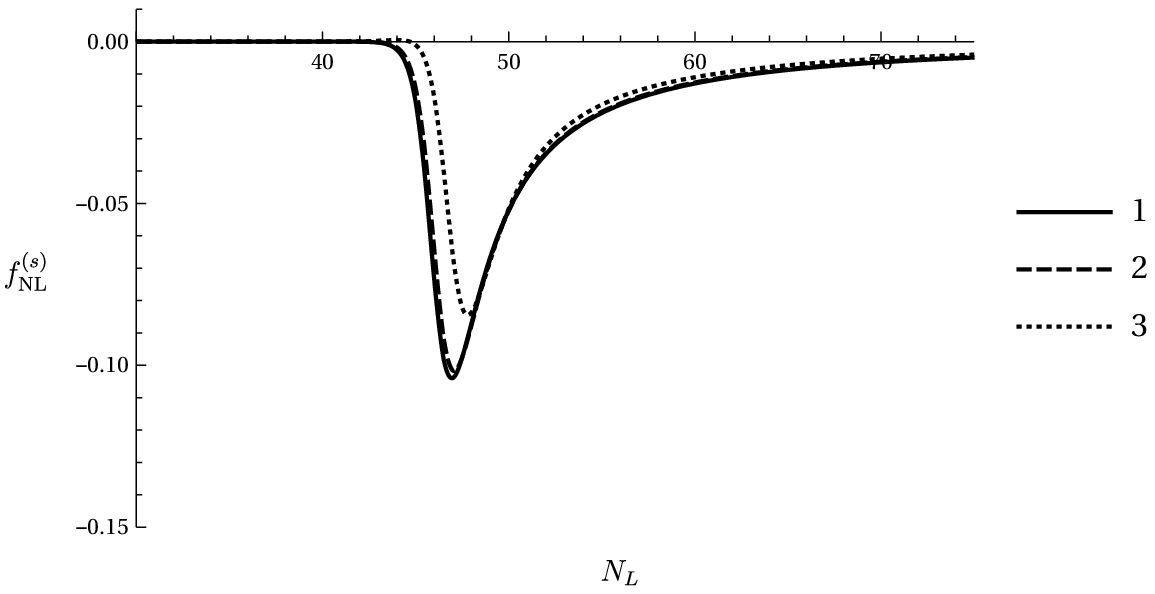}
    \caption{Reduced bispectrum for backward formulation} 
    \label{fnlbvl2}
    \end{subfigure} 
	\caption{
In these plots, lines 1, 2, and 3 correspond to the cases $ \lm_2$ = 0, 0.01, and 0.1, respectively.
For all lines, $ \lm_1 =1, p=1 $.
} 
\label{fig:varyl2}
\end{figure}

\begin{figure}[!h]
\centering
	\begin{minipage}[t]{0.45\textwidth}
        \includegraphics[width=\textwidth]{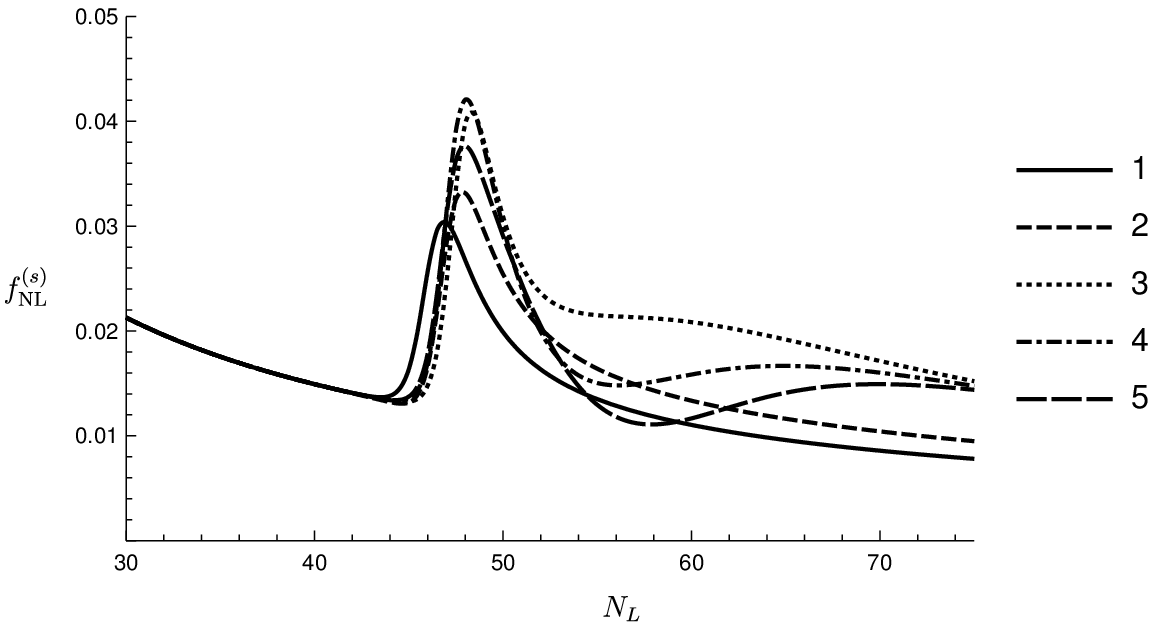}
        \subcaption{Reduced bispectrum for forward formulation} 
        \label{fnlfvp}
    \end{minipage}
    \begin{minipage}[t]{0.45\textwidth}
        \includegraphics[width=\textwidth]{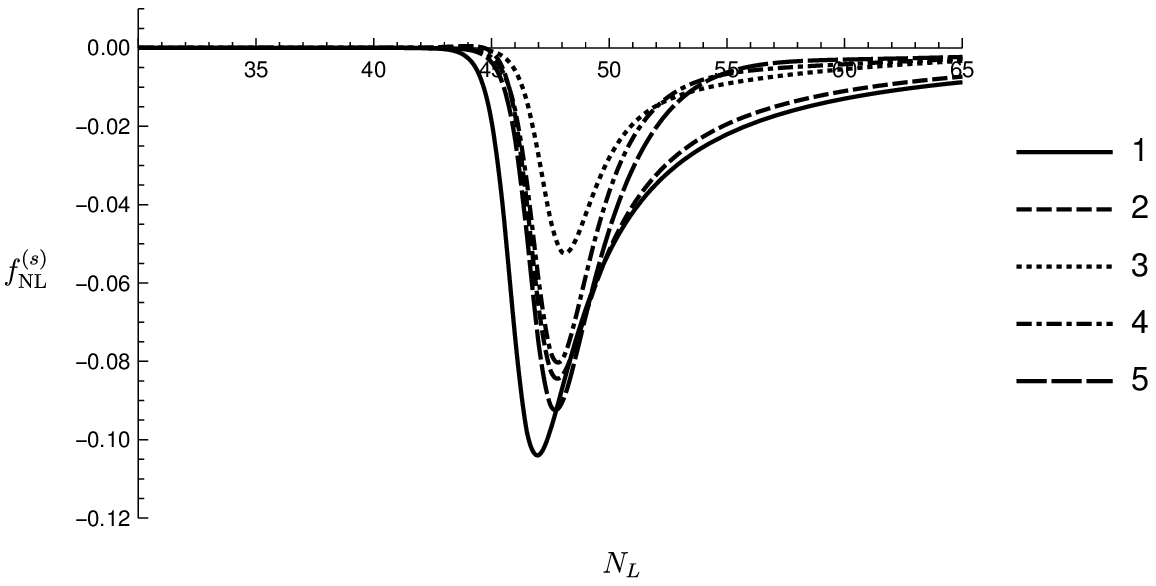}
        \subcaption{Reduced bispectrum for backward formulation} 
        \label{fnlbvp}
    \end{minipage} 
\caption{
In these plots, lines 1, 2, 3, 4, and 5 correspond to the cases where 
$(\lm_2, p) = (0,1), (0.1, 1), (0.1,2), (0.01,3), (0.001, 4)$, respectively.
For all lines, $\lm_1 = 1$.
} 
\label{fig:fnlvp}
\end{figure}

Plots of $\fnl^\sz$ for various values of parameters $p, \lm_1, \lm_2$ are shown in figures (\ref{fig:varyl1}) - (\ref{fnlfvp}).
For all plots, we  set $m_\phi = 9 \times 10^6$GeV, $m_\phi / m_\chi =9$,
$N_S = N_L -7$, and $N_L$ is set to be zero at the end of inflation.
For a given value of $p$, $\lm_2$ are chosen such that $\partial G / \partial\phi$ is small enough to make Eq.~(\ref{kg-slow}) valid.
The amplitude of the reduce bispectrum can be altered if $N_L - N_S$ is different from 7,
but the main conclusions in the following discussion are still unchanged.

\begin{figure}
\begin{center}
\includegraphics[width=0.8\textwidth, height=0.8\textwidth]{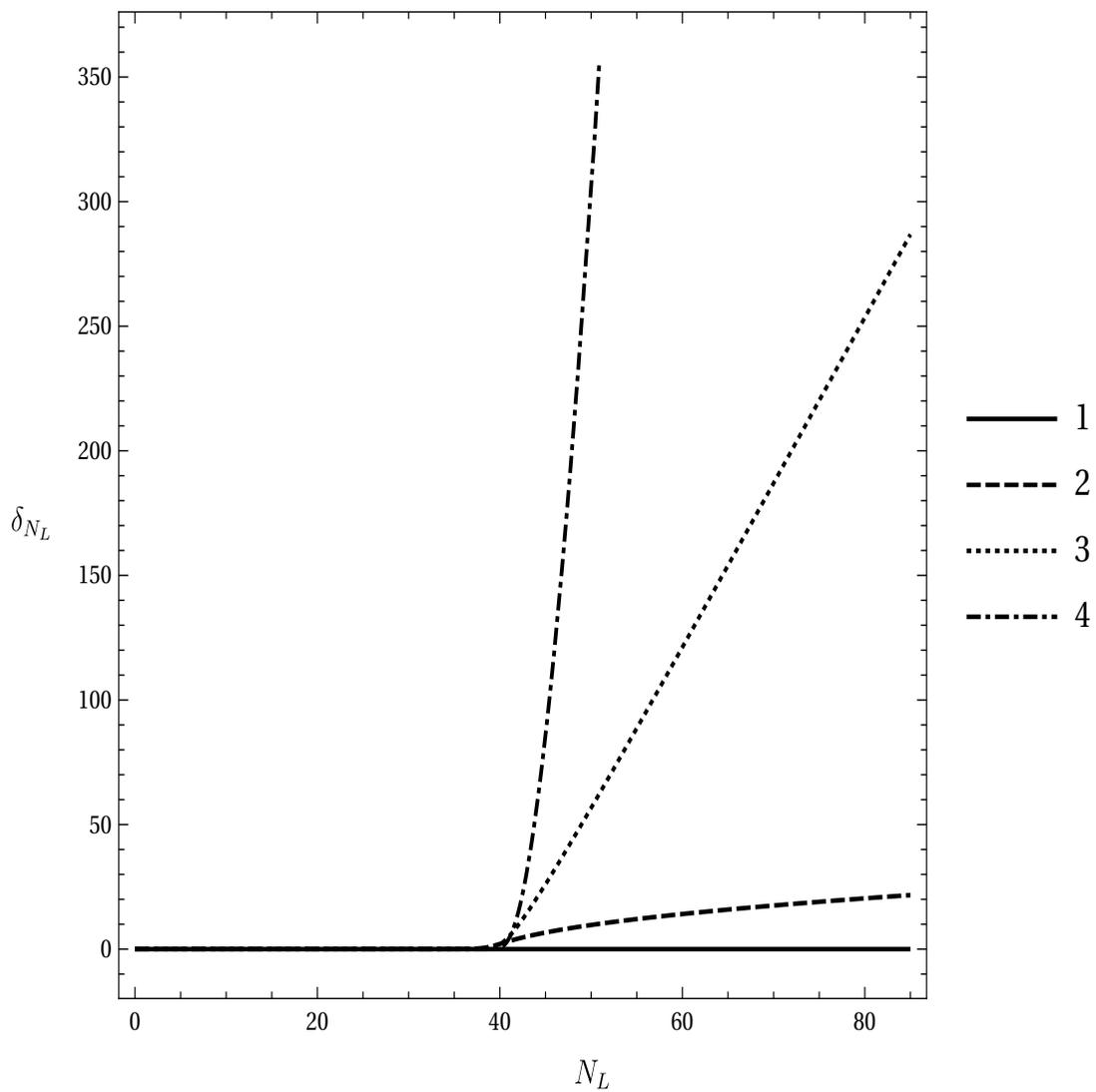}
\caption{\label{fig:dn}
Plots of $\delta_N$  as a function of $N$.
Lines 1, 2, 3 and 4 correspond to the cases where $(\lm_1, \lm_2, p) = (1.5,0,1), (1,0.01,1), (1,0.01,2)$, and $ (1,0.01,3)$,
respectively.
}
\end{center}
 \end{figure}

From figure (\ref{fig:varyl1}), we see that peak position of $\fnl^\sz$ shifts to larger $N_L$ for both the forward and backward cases when $\lm_1$ increases.
This is a consequence of the reduction of initial value of $\phi$ due to the enhancement of $\lm_1$,
which shifts the transition between the $\phi$ dominated and  $\chi$ dominated epoched to the early stage of inflation.
The decreasing of initial value of $\phi$ also suppresses the maximum value of slow-roll parameter $\epsilon$ during the transition between $\phi$ domination and $\chi$ domination.
It follows from Eq.~(\ref{fnl-sq-ni-rep}) that one of the contributions to $\fnl^\sz$ is proportional to $\epsilon$.
According to our numerical check,
the magnitude of $\fnl^\sz$  strongly depends on the magnitude of slow-roll parameter during the transistion stage,
so that the increasing of $\lm_1$ leads to suppression of the peak amplitude of $\fnl^\sz$.
It follows from figures (\ref{fig:varyl2}) -- (\ref{fig:fnlvp}) that
the increasing of $\lm_2$ shifts peak position of the $\fnl^\sz$ to larger $N_L$,
and enhances  peak amplitude of the forward  $\fnl^\sz$ but suppress peak amplitude of backward $\fnl^\sz$.
These effects of $\lm_2$ on the peak amplitude of $\fnl^\sz$ are consequences of positive contributions from the terms that are proportional to $\partial G / \partial\phi$ and $\delta_N$ in Eqs.~(\ref{fnlbak}) and (\ref{pnsrep}).
The influences of $\lm_2$ on $\fnl^\sz$ are stronger when $p$ increases,
which is a result from enhancement of $\delta_N$ due to an increasing  $p$ as shown in figure (\ref{fig:dn}).
Since  $\delta_N$ increases at larg $N_L$ when $\epsilon$ is still small,
this quantity and also $\partial G / \partial\phi$ can influence $\fnl^\sz$ at larg $N_L$ for the cases where $p > 1$.
As a result, the shape of $\fnl^\sz$ is modified when $p>1$.

\begin{figure}[!h]
\centering
		\begin{minipage}[b]{0.45\textwidth}
    \includegraphics[width=\textwidth]{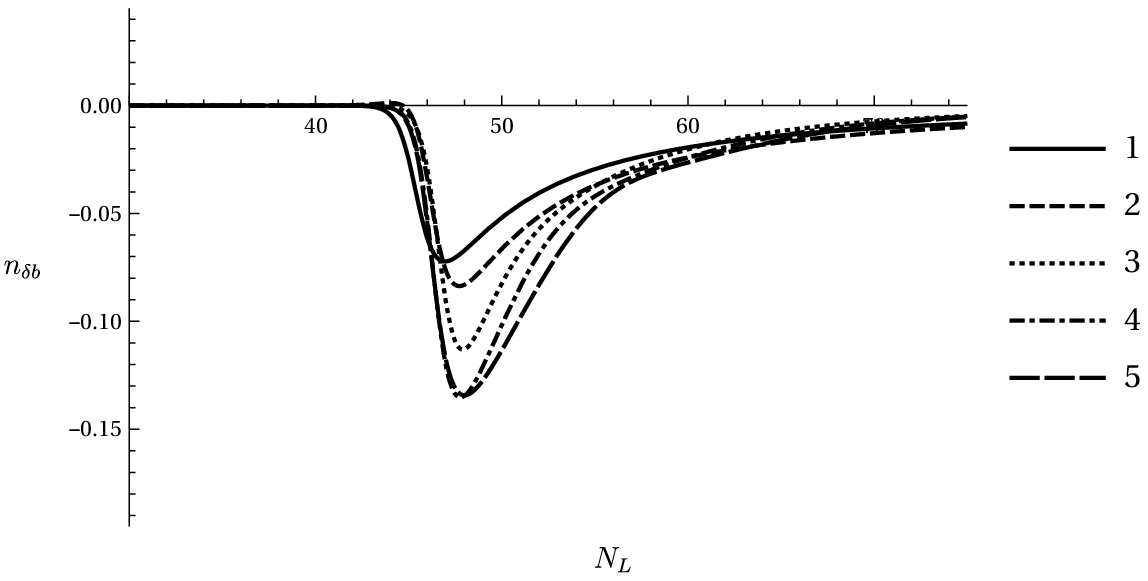}
	\subcaption{spectral index of the halo bias for forward formulation} 
	\label{ndbf}
    \end{minipage}
    \begin{minipage}[b]{0.45\textwidth}
    \includegraphics[width=\textwidth]{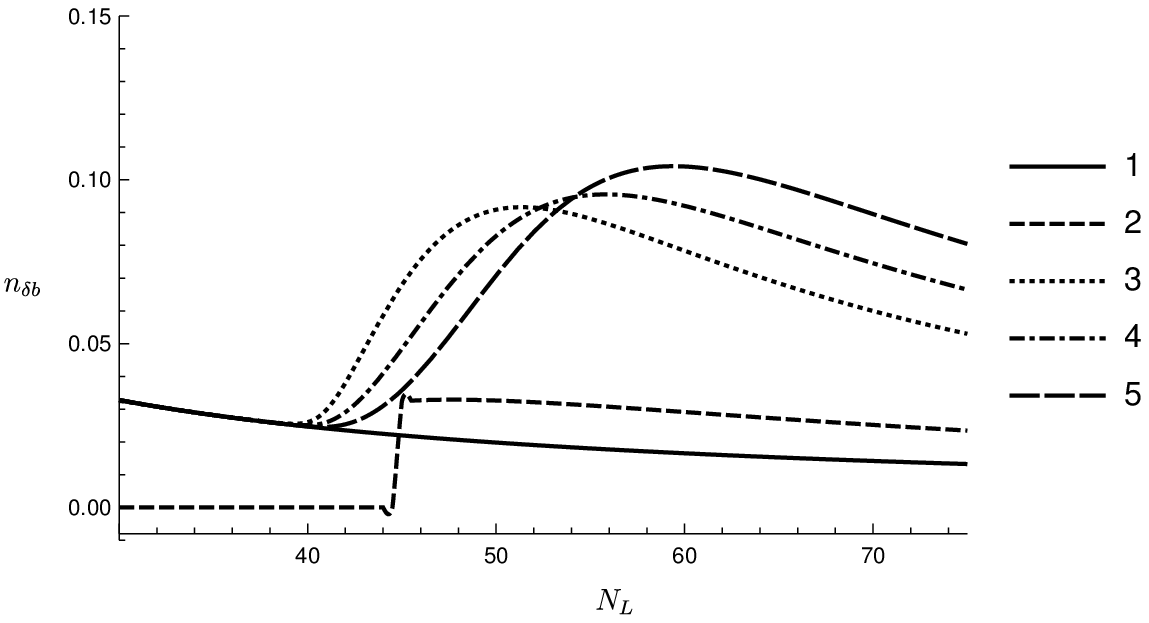}
	\subcaption{spectral index of the halo bias for backward formulation} 
    \end{minipage}    
	\label{ndbb}
\caption{
In these plots, lines 1, 2, 3, 4, and 5 correspond to the cases where 
$(\lm_2, p) = (0,1), (0.1, 1), (0.1,2), (0.01,3), (0.001, 4)$, respectively.
For all lines, $\lm_1 = 1$.
} 
\label{fig:ndb}
\end{figure}

The shape of $\fnl$ is characterized by the spectral indices which describe how it depends on scales or wavenumber.
For squeezed bispectrum, the dependence of $k_1^3 B_{\zeta}(k_1,k_2,k_2)$ on the squeezed wavenumber $k_1$ influences spectral index of the halo bias, $n_{\delta b}$ \cite{halo} through the relation $n_{\delta b} \equiv n_{\rm sz} - (n_s-1)$ \cite{Kenton:15, Dias:2013rla}.
Here, $n_s$ is the spectral index of the power spectrum at $k_1$,
and $n_{\rm sz}$ is the tilt of the squeezed bispectrum with respect to $k_1$.
The spectral index $n_{\rm sz}$ can be computed by
\be
n_{\rm sz} = \frac{\partial \ln \(k_1^3B_{\zeta}(k_1,k_2,k_2)\)}{\partial \ln k_1}
= \frac 1{k_1^3 B_{\zeta}}
\frac{\partial \varphi_L^I}{\partial N} \frac{\partial}{\partial \varphi_L^I}\(k_1^3 B_\zeta\)\,.
\label{nsz}
\ee
The squeezed bispectrum $B_\zeta(k_1, k_2, k_2)$ can be computed by substituting Eqs.~(\ref{p-field}) and (\ref{fnl-sq-ni-rep}) into Eq.~(\ref{fnl-sq}) .
Inserting  the obtained $B_\zeta$ in the above equation,
we get
\ba
n_{\rm sz}
&=& - \frac{\[\(2 N_{K' K} N^{K'} + N_{J'} N_{K'} G^{J' K'}_{,K}\) W_S + N_{I'} N^{I'} W_{,K}\]_S}
{N_I|_L G^{IJ}(\phi_L)\[\(2 N_{K' K} N^{K'} + N_{J'} N_{K'} G^{J' K'}_{,K}\) W_S + N_{I'} N^{I'} W_{,K}\]_S \Gamma_J^K} 
\times
\nonumber\\
&&
\[2 \epsilon|_L N^J|_L\Gamma_J^K + \left. \frac{W^{,M}}{W}\right|_L \(N^J|_L \frac{\partial \Gamma_J^K}{\partial \varphi^M_L} + \Gamma_J^K N^J_M|_L\)\]\,,
\label{nszrep}
\ea
where
\be
\Gamma_J^K \equiv \left. \frac{\partial \varphi^K_S}{\partial \varphi^J}\right|_L.
\ee
The spectral index of the power spectrum at $k_1$ is computed from the relation
$ n_s - 1 = \frac{\partial \ln (k_1^3 P_\zeta(k_1)) }{ \partial \ln k_1 }$ which is obtained from Eq.~(\ref{pindex}) by evaluating this equation at $N_L$ instead $N_S$.
The explicit expressions for $n_{\rm sz}$ and $n_{\delta b}$ in terms of the fields $\phi$ and $\chi$ as well as their potential and $G(\phi)$ can be computed using the expressions in the appendices (\ref{appena}) and (\ref{appenc}).
Nevertheless, their expressions are rather complicated.
Hence, we do not present there expressions here,
but  plot them for varius values of parameters in figure (\ref{fig:ndb}).
From the figure, we see that the  maximum value of $|n_{\delta b}|$ for the case $p > 1$ can be 2 -- 4 times larger than that for the case of flat field-space metric,
which could be observable in future servays \cite{Euclid, LSST}.

\section{Conclusions}
\label{conclusions}

In this work, we investigate effects of the curved field-space metric in multi-field inflationary model on squeezed bispectrum.
Based on the slow-roll approximation and the assumption that the slope of the field-space metric is nearly flat,
we compute reduced bispectrum in squeezed limit and corresponding spectral index of the halo bias for both the backward and forward formulations using $\delta N$ formalism.
According to our analytic expressions for the reduced bispectrum,
the effects of the nearly flat field-space metric on reduced bispectrum depend on $\partial G / \partial \phi$, i.e., slope of the field  metric, and $\delta_N$ defined in Eq.~(\ref{dn}).
We find that the amplitude and shape of the reduced bispectrum can be altered compared with the flat field-space metric by these two quantities.
The modification  of the reduced bispectrum due to these two quantities leads to spectral index of the halo bias with amplitude larger than that for the flat field-space metric model by factor  2 -- 4.
This feature of the curved field-space metric could leave observational imprints in future galaxy surveys.

\section*{Acknowledgements}
SJ was supported by full scholarship of Naresuan University for master degree student.

\appendix
\section{Calculations of $N_I, N_{IJ}$ and $\partial \varphi^J_S / \partial\varphi^K |_L$}
\label{appena}

\subsection{$N_I$}
In order to compute $N_I$,
we set $t_1$ and $t_2$ in Eq.~(\ref{def-n}) to be times at which the specific  perturbations modes are on the spatially flat and uniform density hypersurfaces respectively.
Thus we have
\be
N = \int_{\phi_u}^{\phi_*} \frac{U(\phi)}{U_\phi(\phi)} d\phi
+ \int_{\chi_u}^{\chi_*} G(\phi(\chi)) \frac{V(\chi)}{V_\chi(\chi)} d\chi\,,
\label{def-n1}
\ee
where subscripts ${}_*$ and ${}_u$ denote evaluation at the time when perturbations are on the spatially flat and uniform density hypersurfaces.
Differentiating the above equation with respect to $\phi_*$ and $\chi_*$,
we get
\ba
N_\phi &\equiv& \frac{\partial N}{\partial \phi_*}
= \frac{U_*}{U_{\phs}} - \frac{\partial \phi_u}{\partial \phi_*} \frac{U_u}{U_{\phu}}
- \frac{\partial \chi_u}{\partial \phi_*} G(\phi_*) \frac{V_u}{V_{\chu}}
+  \int_{\chi_u}^{\chi_*} \frac{\partial G(\phi(\chi))}{\partial \phi_*} \frac{V(\chi)}{V_\chi(\chi)} d\chi\,,
\label{dnps}\\
N_\chi &\equiv& \frac{\partial N}{\partial \chi_*}
= G(\phi_*) \frac{V_*}{V_{\chs}} - \frac{\partial \chi_u}{\partial \chi_*} G(\phi_u) \frac{V_u}{V_{\chu}}
- \frac{\partial \phi_u}{\partial \chi_*} \frac{U_u}{U_{\phu}}
+ \int_{\chi_u}^{\chi_*} \frac{\partial G(\phi(\chi))}{\partial \chi_*} \frac{V(\chi)}{V_\chi(\chi)} d\chi\,.
\label{dncs}
\ea
The derivative of fields at $t = t_u$  with respect to fields at $t = t_*$
can be computed from Eq.~(\ref{constr}) and the condition $\de\rho = 0$ on uniform density hypersurfaces.
Setting $(\phi_1, \chi_1)$ and $(\phi_2, \chi_2)$ in Eq.~(\ref{constr}) to be $(\phi_*, \chi_*)$ and $(\phi_u, \chi_u)$, and differentiating the result with respect to $\phi_*$ and $\chi_*$, we get
\ba
0 &=& \frac 1{ U_{\phs} G(\phi_*)} - \frac{\partial \phi_u}{\partial \phi_*} \frac 1{U_{\phu} G(\phi_u)}
 + \frac{\partial \chi_u}{\partial \phi_*} \frac 1{V_{\chu}}
\label{dconstr1}\\
0 &=& \frac{\partial \phi_u}{\partial \chi_*} \frac 1{U_{\phu} G(\phi_u)} 
+ \frac 1{V_{\chs}} - \frac{\partial \chi_u}{\partial \chi_*} \frac 1{V_{\chu}}\,.
\label{dconstr2}
\ea
On the uniform density hypersurfaces, there are no perturbations in energy density $\de\rho = 0$,
so that if the slow roll approximation is assumed, we have $\de \rho \simeq \de W = U_{\phi u} \de\phi + U_{\chi u}\de\chi$
and therefore
\ba
0 &=& U_{\phu} \frac{\partial \phi_u}{\partial \phi_*} + V_{\chu} \frac{\partial \chi_u}{\partial \phi_*}\,,
\nonumber\\
0 &=& U_{\phu} \frac{\partial \phi_u}{\partial \chi_*} + V_{\chu} \frac{\partial \chi_u}{\partial \chi_*}\,.
\ea
Solving the above four equations,
we obtain
\ba
\frac{\pa \phi_u}{\pa \phi_*} 
&=& \frac{G\left(\phi _u\right) U_{\phi _u} V_{\chi _u}^2}{G\left(\phi _*\right) U_{\phi _*} V_{\chi _u}^2+G\left(\phi _*\right) U_{\phi _*} G\left(\phi _u\right) U_{\phi _u}^2}\,,
\quad
\frac{\pa \chi_u}{\pa \phi_*} 
=  -\frac{G\left(\phi _u\right) U_{\phi _u}^2 V_{\chi _u}}{G\left(\phi _*\right) U_{\phi _*} V_{\chi _u}^2+G\left(\phi _*\right) U_{\phi _*} G\left(\phi _u\right) U_{\phi _u}^2}\,,
\nonumber\\
\frac{\pa \phi_u}{\pa \chi_*} 
&=& -\frac{G\left(\phi _u\right) U_{\phi _u} V_{\chi _u}^2}{V_{\chi _*} G\left(\phi _u\right) U_{\phi _u}^2+V_{\chi _*} V_{\chi _u}^2}\,,
\quad
\frac{\pa \chi_u}{\pa \chi_*} 
=  \frac{G\left(\phi _u\right) U_{\phi _u}^2 V_{\chi _u}}{V_{\chi _*} G\left(\phi _u\right) U_{\phi _u}^2+V_{\chi _*} V_{\chi _u}^2}\,.
\label{dfu2dfs}
\ea
In order to evaluate the integration terms in Eqs.~(\ref{dnps}) and (\ref{dncs}),
we express the field-space metric $G$ in terms of $\chi$ by inserting Eq.~(\ref{pofc}) into Eq.~(\ref{gij}) as
\be
G(\phi(\chi)) = \frac{\lm_1}{1 - \lm_2 \fp(\phi_1) (\chi / \chi_1)^{r p}}\,.
\label{ch2g}
\ee
Setting $(\phi_1, \chi_1) = (\phi_*, \chi_*)$,
and differentiating this equation with respect to $\phi_*$ and $\chi_*$ ,
we respectively get
\ba
\frac{\pa G(\phi(\chi))}{\partial \phi_*} 
&=& \frac{\pa \fp(\phi_*)}{\partial \phi_*} \frac{\lm_1 \lm_2 \fp(\phi_*) (\chi / \chi_*)^{r p}}{\(1 - \lm_2 \fp(\phi_*) (\chi / \chi_*)^{r p}\)^2}
= \frac 1{\lm_1 \fp(\phi_*)}  \frac{\pa \fp(\phi_*)}{\partial \phi_*} G(\phi) \phi^p\,,
\\
\frac{\pa G(\phi(\chi))}{\partial \chi_*} 
&=& - \frac1{\chi_*}\frac{\lm_1 r p \lm_2 \fp(\phi_*) (\chi / \chi_*)^{r p}}{\(1 - \lm_2 \fp(\phi_*) (\chi / \chi_*)^{r p}\)^2}
= - \frac{r p \lm_2}{\lm_1 \chi_*} G(\phi) \phi^p\,.
\ea
Inserting the above relation into the integration terms in Eqs.~(\ref{dnps}) and (\ref{dncs})
and performing suitable integration by parts,
we can write the integration terms in terms of the number of e-folding given Eq.~(\ref{def-n1}) and obtain
\ba
N_\phi &=& \frac{G_* U_* V_{\chi _u}^2+G_* U_* G_u U_{\phi _u}^2-G_u U_u V_{\chi _u}^2+G_u^2 V_u U_{\phi _u}^2}
{G_* U_{\phi _*} V_{\chi _u}^2+G_* G_u U_{\phi _*} U_{\phi _u}^2}
+ \frac{\lambda _1 \delta _{N_*}}{2 r G_* \phi _*}\,,
\label{dndphi}\\
N_\chi &=&\frac{G_* V_*+\frac{G_u \left(U_u V_{\chi _u}^2-G_u V_u U_{\phi _u}^2\right)}{G_u U_{\phi _u}^2+V_{\chi _u}^2}}{V_{\chi _*}}
-\frac{\delta _{N_*}}{2 \chi _*}\,,
\label{dndchi}
\ea
where we have used $d \fp / d \phi_s = p \lm_1 \fp / (\phi_* G_*)$ and
\be
\delta_{N_*} \equiv
 - 4 N_* + \phi_*^2 - \phi_u^2 + G\(\phi_*\) \chi_*^2 - G\(\phi_u\) \chi_u^2\,.
\label{dn}
\ee
The number of e-folding $N_*$ in the above equation is given by Eq.~(\ref{def-n1}).
It is clear that $\delta_N = 0$ when the curvature in field-space disappears, i.e., $\lm_2 = 0$.
From figure (\ref{fig:1}),
we see that   $\phi \to 0$ at the end of inflation.
Since we are interested to evaluate $\fnl$ at the end of inflation,
we set $t_u$ to be a time at the end of inflation, so that we have $\phi_u \sim 0$
and therefore the above equations become
\be
N_\phi = \frac{\phi _*}{2} + \frac{\lambda _1 \delta _{N_*}}{2 r G_* \phi _*}\,,
\quad
N_\chi = \frac{G_* \chi _*}{2} -\frac{\delta_{N_*}}{2 \chi _*}\,.
\label{nisim}
\ee
where we have insert the expressions for $U$ and $V$ from Eq.~(\ref{pot}) into the above equation.

\subsection{$N_{IJ}$}

Since we are interested in the case $\phi_u \sim 0$,
$N_{IJ}$ can be computed by differentiating Eq.~(\ref{nisim}) with respect to $\phi_*$ and $\chi_*$ and the results are
\be
N_{IJ} = 
\( \begin{array}{cc}
N_{\phi\phi} & N_{\phi\chi} \\
N_{\chi\phi}& N_{\chi\chi}\\
                 \end{array}\)
= \( \begin{array}{cc}
\frac{1}{2} & 0 \\
0 & \frac{G_*}{2} \\
                 \end{array}\)
+ \(\begin{array}{cc}
- \lm_1 \frac{\delta_{N_*} \left(G_* r +2 \lambda _1\right)+G_{\phi _*} r \phi _* \left(\delta_{N_*}-G_* \chi _*^2\right)}{2 G_*^2 r^2 \phi _*^2} 
& \frac{\delta_{N_*} \lambda _1}{G_* r \phi _* \chi _*} \\
\frac{\delta_{N_*} \lambda _1}{G_* r \phi _* \chi _*}
& -\frac{\delta_{N_*}}{2 \chi _*^2} \\
\end{array}\)\,,
\label{nijsim}
\ee
where $G_{\phi_*} \equiv \left. \partial G(\phi) / \partial\phi \right|_{*}$.

\subsection{$\partial \varphi^J_S / \partial\varphi^K|_L$}

We first set $(\phi_1, \chi_1)$ and $(\phi_2, \chi_2)$ in Eq.~(\ref{constr}) to $(\phi_L, \chi_L)$ and $(\phi_S, \chi_S)$.
Differentiating  the result with respect to $\phi_L$ and $\chi_L$,
we respectively get
\ba
0 &=& \frac 1{ U_{\phi_L} G(\phi_L)} - \frac{\partial \phi_S}{\partial \phi_L} \frac 1{U_{\phi_S} G(\phi_S)}
 + \frac{\partial \chi_S}{\partial \phi_L} \frac 1{V_{\chi_S}}
\label{constr1}\\
0 &=& \frac{\partial \phi_S}{\partial \chi_L} \frac 1{U_{\phi_S} G(\phi_S)} 
+ \frac 1{V_{\chi_L}} - \frac{\partial \chi_S}{\partial \chi_L} \frac 1{V_{\chi_S}}\,.
\label{constr2}
\ea
In order to solve the above equations for  $\partial \varphi^J_S / \partial\varphi^K_L$,
we need two more equations of $\partial \varphi^J_S / \partial\varphi^K_L$.
The required equations can be obtained by differentiate the equation for the number of e-folding with respect to $\phi_L$ and $\chi_L$.
However, there are two choices of the specification of a number of e-folding at which the short wavelength perturbation mode exits the Hubble radius.
We consider each specification separately in the following sections.

\subsubsection{Forward formulation}

For the forward formulation,
$N_S$ is specified from $N_L$ such that $N_L - N_S$ is constant against variations of $\phi_L$ and $\chi_L$.
Here, $N_L$ and $N_S$ are the number of e-folding realised backwards in time from the end of inflation to times at which the long and short wavelength exit Hubble radius respectively.
Hence, we have
\be
N_L - N_S = \int_{\phi_S}^{\phi_L} \frac{U(\phi)}{U_\phi(\phi)} d\phi
+ \int_{\chi_S}^{\chi_L} G(\phi(\chi)) \frac{V(\chi)}{V_\chi(\chi)} d\chi = \mbox{constant}\,.
\label{nabf}
\ee
Differentiating the above equations with respect to $\phi_L$ and $\chi_L$,
we get 
\ba
0 &=& \frac{U_L}{U_{\phi_L}} - \frac{\partial \phi_S}{\partial \phi_L} \frac{U_S}{U_{\phi_S}}
- \frac{\partial \chi_S}{\partial \phi_L} G_L \frac{V_S}{V_{\chi_S}}
+  \int_{\chi_S}^{\chi_L} \frac{\partial G(\phi(\chi))}{\partial \phi_*} \frac{V(\chi)}{V_\chi(\chi)} d\chi\,,
\label{dnpa}\\
0 &=& G_L \frac{V_L}{V_{\chi_L}} - \frac{\partial \chi_S}{\partial \chi_L} G_S \frac{V_S}{V_{\chi_S}}
- \frac{\partial \phi_S}{\partial \chi_L} \frac{U_S}{U_{\phi_S}}
+ \int_{\chi_S}^{\chi_L} \frac{\partial G(\phi(\chi))}{\partial \chi_*} \frac{V(\chi)}{V_\chi(\chi)} d\chi\,.
\label{dnca}
\ea
Expressing the integrations in the above equations in terms of the number of e-folding
and solving Eq.~(\ref{constr1}), (\ref{constr2}), (\ref{dnpa}) and (\ref{dnca}),
we obtain
\ba
\Gamma^f_{SL} &\equiv&
\(\begin{array}{cc}
\frac{\partial \phi_S}{\partial \phi_L} & \frac{\partial \phi_S}{\partial \chi_L} \\
\frac{\partial \chi_S}{\partial \phi_L} & \frac{\partial \chi_S}{\partial \chi_L} \\
\end{array}\)
\label{gf}\\
&=&
\(\begin{array}{cc}
\frac{U_{\phi _S} \left(G_L U_L+G_S V_S\right)}{G_L U_{\phi _L} W_S} & \frac{U_{\phi _S} \left(G_L V_L-G_S V_S\right)}{V_{\chi _L} W_S} \\
\frac{V_{\chi _S} \(G_L U_L -G_S U_S\)}{G_L G_S U_{\phi _L} W_S} & \frac{V_{\chi _S}\left(G_S U_S+G_L V_L\right)}{G_S V_{\chi _L} W_S} \\
                              \end{array}\)
+ 
\(\begin{array}{cc}
\frac{U_{\phi _S}}{W_S} I_1 & \frac{U_{\phi _S}}{W_S} + I_2\\
\frac{V_{\chi _S}}{G_S W_S} I_1 & \frac{V_{\chi _S}}{G_S W_S} + I_2\\
               \end{array}\)\,,
\nonumber
\ea
where 
\ba
I_1 &\equiv& \frac{\lambda _1 \delta _{N_{LS}}}{2 r G_L \phi _L}\,,
\nonumber\\
I_2 &\equiv& - \frac{\delta _{N_{LS}}}{2 \chi _L}\,,
\nonumber\\
\delta_{N_{LS}} &\equiv &
\delta_{N_L} - \delta_{N_S} =
- 4 N_{LS} + \phi_L^2 - \phi_S^2 + G\(\phi_L\) \chi_L^2 - G\(\phi_S\) \chi_S^2\,,
\label{dnBA}
\ea
where  $N_{LS} \equiv N_L - N_S$ which is given in Eq.~(\ref{nabf}),
while $\delta_{N_L}$ and $\delta_{N_S}$ are given by Eq.~(\ref{dn}) with $(\phi_*, \chi_*) = (\phi_L, \chi_L)$
and $(\phi_*, \chi_*) = (\phi_S, \chi_S)$ respectively.
Substituting the expressions for the potentials from Eq.~(\ref{pot}) into Eq.~(\ref{gf}),
we get
\ba
\Gamma^f_{SL} 
=
\(\begin{array}{cc}
\frac{1}{2W_S} \left(\phi _L \phi _S m_{\phi }^2+\frac{G_S m_{\chi }^2 \phi _S \chi _S^2}{G_L \phi _L}+\frac{m_{\chi }^2 \delta_{N_{LS}}}{G_S}\right)
& -\frac{m_{\phi }^2 \phi _S }{2 \chi _L \chi _S W_S}\left(G_S \chi _S^3-G_L \chi _L^2 \chi _S+\delta_{N_{LS}} \chi _L\right)
\\
\frac{1}{2W_S} m_{\chi }^2 \chi _S \left(\frac{\delta_{N_{LS}} m_{\chi }^2}{m_{\phi }^2 G_S^2 \phi _S} + \frac{\phi _L}{G_S} -\frac{\phi _S^2}{G_L \phi _L}\right)
& \frac{-\delta_{N_{LS}} \chi _L m_{\chi }^2+G_L \chi _L^2 \chi _S m_{\chi }^2+G_S m_{\phi }^2 \phi _S^2 \chi _S}{2 G_S \chi _L W_S} \\
\end{array}\)\,.
\label{gfrep}
\ea
When the expressions for the potentials are inserted into Eq.~(\ref{dnBA}),
we have
\be
\delta_{N_{LS}} = 
- 2\int_{\chi_S}^{\chi_L} G(\phi(\chi)) \chi d\chi + G\(\phi_L\) \chi_L^2 - G\(\phi_S\) \chi_S^2\,.
\label{dnBA-r}
\ee
Using Eq.~(\ref{n-hyper}), the above equation can be written as
\be
\delta_{N_{LS}} = 
\lm_1 \chi_S^2 {\rm Hy}(\chi_S) - \lm_1 \chi_L^2 {\rm Hy}(\chi_L)
+ G\(\phi_L\) \chi_L^2 - G\(\phi_S\) \chi_S^2\,,
\label{dnBA-hy}
\ee
where 
\be
{\rm Hy}(\chi) \equiv
{}_2F_1\( 1 , \frac{2}{p r} ; 1 + \frac{2}{p r} ; \( \frac{\chi}{\chi_i} \)^{p r} \lm_2 \fp(\phi_i)\).
\label{hy-red}
\ee

\subsubsection{Backward formulation}

In the backward formulation,
$N_S$ and $N_L$ are defined as the number of e-folding 
realised backwards in time from the end of inflation to times at which the short and long wevlength perturbations exit horizon respectively.
In this formulation, $N_S$is fixed against variations of $\phi_L$ and $\chi_L$,
but $N_L - N_S$ is not necessary constant different from the case of forward formulation.
For this case, the additional equations for $\partial \varphi^J_S / \partial\varphi^K_L$ can be obtained by differentiating equation
\be
N_S = \int_{\phi_u}^{\phi_S} \frac{U(\phi)}{U_\phi(\phi)} d\phi
+ \int_{\chi_u}^{\chi_S} G(\phi(\chi)) \frac{V(\chi)}{V_\chi(\chi)} d\chi\,,
\label{nbf}
\ee
with respect to $\phi_L$ and $\chi_L$.
The differentiation gives two equations describing relations among $\pa \varphi^I_S / \pa \varphi^J_L$ and $\pa\varphi^I_u / \pa\varphi^J_L$.
The expressions for $\pa\varphi^I_u / \pa\varphi^J_L$ can be computed using the same approach as for Eq.~(\ref{dfu2dfs}),
and the results take similar form as in Eq.~(\ref{dfu2dfs}) with the replacement of evaluation at $N_*$ by evaluation at $N_L$.
Inserting these results into the relations among $\pa \varphi^I_S / \pa \varphi^J_L$ and $\pa\varphi^I_u / \pa\varphi^J_L$ obtained from the differentiate of Eq.~(\ref{nbf}) with respect to $\phi_L$ and $\chi_L$,
we get two relations for $\pa \varphi^I_S / \pa \varphi^J_L$.
Solving these two relations together with Eqs.~(\ref{constr1}) and (\ref{constr2}),
we obtain
\ba
\frac{\partial \phi_S}{\partial \phi_L}
&=&
\frac{U_{\phi _S} \(G_u \(G_S V_S-G_u V_u\) U_{\phi _u}^2+\(G_u U_u+G_S V_S\) V_{\chi _u}^2\)}{G_L U_{\phi _L} W_S \(G_u U_{\phi _u}^2+V_{\chi _u}^2\)} 
\nonumber\\
&& -\frac{U_{\phi _S}}{W_S} \frac{\lambda _1 \delta _{N_S}}{2 r G_L \phi_L}\,,
\\
\frac{\partial \phi_S}{\partial \chi_L}
&=&
-\frac{U_{\phi _S} \(G_u \(G_S V_S-G_u V_u\) U_{\phi _u}^2+\(G_u U_u+G_S V_S\) V_{\chi _u}^2\)}{V_{\chi _L} W_S \(G_u U_{\phi _u}^2+V_{\chi _u}^2\)}
\nonumber\\
&& \frac{U_{\phi _S}}{W_S} \frac{\delta _{N_S}}{2 \chi _L}\,,
\\
\frac{\partial \chi_S}{\partial \phi_L}
&=&
-\frac{V_{\chi _S} \(G_S U_S \(G_u U_{\phi _u}^2+V_{\chi _u}^2\)+G_u \(G_u U_{\phi _u}^2 V_u-U_u V_{\chi _u}^2\)\)}{G_L G_S U_{\phi _L} W_S \(G_u U_{\phi _u}^2+V_{\chi _u}^2\)} 
\nonumber\\
&& -\frac{V_{\chi _S}}{G_S W_S} \frac{\lambda _1 \delta _{N_S}}{2 r G_L \phi_L}\,,
\\
\frac{\partial \chi_S}{\partial \chi_L} 
&=&
\frac{V_{\chi _S} \(G_S U_S \(G_u U_{\phi _u}^2+V_{\chi _u}^2\)+G_u \(G_u U_{\phi _u}^2 V_u-U_u V_{\chi _u}^2\)\)}{G_S W_S V_{\chi _L} \(G_u U_{\phi _u}^2+V_{\chi _u}^2\)}
\nonumber\\
&& -\frac{V_{\chi _S}}{G_S W_S} \frac{\delta _{N_S}}{2 \chi _L}\,,
\ea
where
\be
\delta_{N_S} \equiv
- 4 N_S + \phi_S^2 - \phi_u^2 + G\(\phi_S\) \chi_S^2 - G\(\phi_u\) \chi_u^2\,,
\label{delnb}
\ee
and $N_S$ is given by Eq.~(\ref{nbf}).
Using $\phi_u \sim 0$ and the expressions for the potentials from Eq.~(\ref{pot}),
we get $U_u \sim U_{\phi_u} \sim 0$ and consequently the above equations give
\ba
\Gamma^b_{SL} &\equiv&
\(\begin{array}{cc}
\frac{\partial \phi_S}{\partial \phi_L} & \frac{\partial \phi_S}{\partial \chi_L} \\
\frac{\partial \chi_S}{\partial \phi_L} & \frac{\partial \chi_S}{\partial \chi_L} \\
\end{array}\)
\label{gb1}\\
&=&
\(\begin{array}{cc}
\frac{G_S U_{\phi _S} V_S}{G_L U_{\phi _L}W_S} & -\frac{G_S U_{\phi _S} V_S}{V_{\chi _L} W_S}\\
-\frac{U_S V_{\chi _S}}{G_L U_{\phi _L}W_S} & \frac{U_S V_{\chi _S}}{V_{\chi _L}W_S} \\
\end{array}\)
+
\(\begin{array}{cc}
-\frac{U_{\phi _S}}{W_S} \frac{\lambda _1 \delta _{N_S}}{2 r G_L \phi_L} & \frac{U_{\phi _S}}{W_S} \frac{\delta _{N_S}}{2 \chi _L} \\
-\frac{V_{\chi _S}}{G_S W_S} \frac{\lambda _1 \delta _{N_S}}{2 r G_L \phi_L} 
& 
\frac{V_{\chi _S}}{G_S W_S} \frac{\delta _{N_S}}{2 \chi _L}\\
\end{array}\)\,.
\nonumber
\ea
Substituting the expressions for the potentials from Eq.~(\ref{pot}) into the above equation,
we obtain
\ba
\Gamma^b_{SL} 
&=&
\(\begin{array}{cc}
\frac{\phi_S m_\chi^2}{2 G_L \phi_LW_S} \(G_S \chi_S^2 -\delta _{N_S}\)
& 
-\frac{\phi_S}{2 \chi_L W_S}\(G_S m_\chi^2 \chi_S^2 - m_\phi^2 \delta _{N_S}\)
\\
- \frac{\chi_S m_\chi^2}{2 m_\phi^2 \phi_L G_L W_S}\(m_\phi^2 \phi_S^2 + \frac{m_\chi^2}{G_S} \delta _{N_S}\)
& 
\frac{\chi_S}{2 \chi_L W_S}\(m_\phi^2 \phi_S^2 + \frac{m_\chi^2}{G_S}\delta _{N_S} \)
\end{array}\)\,.
\nonumber\\
{} && \label{gbrep}
\ea
When the expressions for the potentials are inserted into Eq.~(\ref{delnb}),
we have
\be
\delta_{N_{S}} = 
- 2\int_{\chi_u}^{\chi_S} G(\phi(\chi)) \chi d\chi + G\(\phi_S\) \chi_S^2 - \lm_1 \chi_u^2\,.
\label{delnb-r}
\ee
Using Eq.~(\ref{n-hyper}), the above equation can be written as
\be
\delta_{N_{S}} = 
G\(\phi_S\) \chi_S^2 - \lm_1 \chi_S^2 {\rm Hy}(\chi_S)\,,
\label{dnB-hy}
\ee
where $\phi_u \ll 1$ has been used.

\section{Coefficients in the expressions for $\fnl^\sz$ and $n_s$}
\label{appenb}

\subsection{$\fnl^\sz$ in backward formulation}
\label{coeffb}

The coefficients in Eq.~(\ref{fnlbak}) are given by
\ba
D_{S_1} &=&
G_L \phi _L^2 m_{\phi }^4+\chi _L^2 m_{\chi }^4\,,
\\
D_{S_2} &=&
G_L \chi _L^2 \phi _L^2 m_{\phi }^2 \left(m_{\chi }^2-G_L m_{\phi }^2\right)\,,
\\
D_{S_3} &=&
2 G_S m_{\phi }^{2}\(m_\phi^2 \phi_S^2 + m_\chi^2 \chi_S^2\)^2 
\[\chi_L^2\left(\delta_{N_L} m_{\chi }^2 + \phi _L^2 G_L m_{\phi }^2\right)^2 +G_L \phi_L^2 m_\phi^4\left(\delta_{N_L} -G_L \chi _L^2\right)^2\] \times
\nonumber\\
&&
\[\chi_S^2\left(\delta_{N_S} m_{\chi }^2 + \phi _S^2 G_S m_{\phi }^2\right)^2 + G_S \phi_S^2m_\phi^4\left(\delta_{N_S} -G_S \chi _S^2\right)^2\]\,,
\\
C_{S_1} &=&
2 G_S^3 \chi _S^4 \phi _S^4 m_{\phi }^{6} \left(G_S \chi _S^2+\phi _S^2\right) \left(G_S m_{\phi }^2-m_{\chi }^2\right)\,,
\\
C_{S_2} &=&
 G_S^3 \chi _S^6 \phi _S^3 m_{\phi }^{4} \left(\chi _S^2 m_{\chi }^2+\phi _S^2 m_{\phi }^2\right) \left(2 m_{\chi }^2-G_S m_{\phi }^2\right)\,,
\\
C_{S_3} &=&
 -2G_S^2 \chi _S^2 \phi _S^2 m_{\phi }^2 \left(\chi _S^2 \phi _S^2 m_{\phi }^2 \left(-6 G_S m_{\chi }^2 m_{\phi }^2+3 G_S^2 m_{\phi }^4+7 m_{\chi }^4\right)+2 \chi _S^4 m_{\chi }^6+2 \phi _S^4 m_{\chi }^2 m_{\phi }^4\right)\,,
\\
C_{S_4} &=&
G_S^2 \chi _S^4 \phi _S m_{\phi }^2 \left(\chi _S^2 m_{\chi }^2+\phi _S^2 m_{\phi }^2\right) \left(\phi _S^2 m_{\phi }^2 \left(3 G_S m_{\phi }^2-4 m_{\chi }^2\right)+2 \chi _S^2 m_{\chi }^4\right)\,,
\\
C_{S_5} &=&
2G_S \left(\chi _S^2 \phi _S^4 m_{\phi }^4 \left(-6 G_S m_{\chi }^2 m_{\phi }^2+3 G_S^2 m_{\phi }^4-2 m_{\chi }^4\right)-\chi _S^6 \left(G_S m_{\chi }^6 m_{\phi }^2+2 m_{\chi }^8\right)-G_S \phi _S^6 m_{\phi }^8\right)\nonumber\\
&&-14G_S\left(\chi _S^4 \phi _S^2 m_{\chi }^6 m_{\phi }^2\right)\,,
\\
C_{S_6} &=&
-G_S \chi _S^2 \phi _S m_{\phi }^2 \left(\chi _S^2 m_{\chi }^2+\phi _S^2 m_{\phi }^2\right) \left(\phi _S^2 m_{\phi }^2 \left(3 G_S m_{\phi }^2-2 m_{\chi }^2\right)+4 \chi _S^2 m_{\chi }^4\right)\,,
\\
C_{S_7} &=&
-2 \left(G_S \phi _S^4 m_{\phi }^6 \left(G_S m_{\phi }^2-m_{\chi }^2\right)+\chi _S^4 \left(m_{\chi }^8-G_S m_{\chi }^6 m_{\phi }^2\right)\right)\,,
\\
C_{S_8} &=&
\phi _S m_{\phi }^2 \left(\chi _S^2 m_{\chi }^2+\phi _S^2 m_{\phi }^2\right) \left(G_S \phi _S^2 m_{\phi }^4+2 \chi _S^2 m_{\chi }^4\right)\,.
\ea

\subsection{$n_s$}
\label{coefns}

The coefficients in Eq.~(\ref{pnsrep}) are given by
\ba
D_n &=&
2 G_S \(m_\phi^2 \phi_S^2 + m_\chi^2 \chi_S^2\)^2
\[\chi_S^2\left(\delta_{N_S} m_{\chi }^2 + \phi _S^2 G_S m_{\phi }^2\right)^2 + G_S \phi_S^2m_\phi^4\left(\delta_{N_S} -G_S \chi _S^2\right)^2\]\,,
\label{dns}\\
C_{n_1} &=&
2 G_S^2 \chi _S^2 \phi _S^2 m_{\phi }^4 \left(2 G_S \chi _S^4 m_{\chi }^4+\chi _S^2 \phi _S^2 \left(G_S m_{\phi }^2+m_{\chi }^2\right)^2+2 G_S \phi _S^4 m_{\phi }^4\right)\,,
\\
C_{n_2} &=&
G_S^2 \chi _S^4 \phi _S^3 m_{\phi }^4 \left(\chi _S^2 m_{\chi }^2+\phi _S^2 m_{\phi }^2\right) \left(2 m_{\chi }^2-G_S m_{\phi }^2\right)\,,
\\
C_{n_3} &=&
4 G_S \chi _S^2 \phi _S^2 m_{\phi }^2 \left(m_{\chi }^2-G_S m_{\phi }^2\right) \left(G_S \phi _S^2 m_{\phi }^4+\chi _S^2 m_{\chi }^4\right)\,,
\\
C_{n_4} &=&
2 G_S \chi _S^2 \phi _S m_{\phi }^2 \left(G_S \chi _S^2 \phi _S^2 m_{\chi }^2 m_{\phi }^4+\phi _S^4 m_{\phi }^4 \left(G_S m_{\phi }^2-m_{\chi }^2\right)+\chi _S^4 m_{\chi }^6\right)\,,
\\
C_{n_5}&=&
2 \left(G_S \phi _S^4 m_{\phi }^6 \left(G_S m_{\phi }^2-m_{\chi }^2\right)+\chi _S^4 \left(m_{\chi }^8-G_S m_{\chi }^6 m_{\phi }^2\right)\right)\,,
\\
C_{n_6} &=&
-\phi _S m_{\phi}^2 \left(\chi _S^2 m_{\chi }^2+\phi _S^2 m_{\phi }^2\right) \left(G_S \phi _S^2 m_{\phi }^4+2 \chi _S^2 m_{\chi }^4\right)\,.
\ea

\section{Derivatives of $\partial \varphi^J_S / \partial\varphi^K |_L$}
\label{appenc}

Derivative of $\partial \varphi^J_S / \partial\varphi^K |_L$ with respect to $\varphi^I_L$ are computed by differentiating Eq.~(\ref{gfrep}) for forward formulation and Eq.~(\ref{gbrep}) for  backward formulation with respect to $\phi_L$ and $\chi_L$.
The differentiation can be straightforwardly performed except for the terms $\delta_{N_{LS}}$ and $\delta_{N_S}$ given in Eqs.~(\ref{dnBA-r}) and (\ref{delnb-r}).
Differentiation of $\delta_{N_{LS}}$ with respect to $\phi_L$ gives
\ba
\frac{\partial \delta_{N_{LS}}}{\partial \phi_L}
&=&
- 2 \frac{\partial }{\partial \phi_L}\int_{\chi_S}^{\chi_L} G(\phi(\chi)) \chi d\chi 
+ 2 G\(\phi_L\) \chi_L\frac{\partial \chi_L}{\phi_L} - 2 G\(\phi_S\) \chi_S \frac{\partial \chi_S}{\phi_L}
\nonumber\\
&&
+  \frac{d G\(\phi_L\)}{d\phi_L} \chi_L^2 - \frac{\partial G\(\phi_S\)}{\partial \phi_L} \chi_S^2,
\nonumber\\
&=&
- 2 \int_{\chi_S}^{\chi_L} \frac{\partial G(\phi(\chi))}{\partial \phi_L} \chi d\chi 
+  G_{\phi_L} \chi_L^2 - \frac{\partial G\(\phi_S\)}{\partial \phi_L} \chi_S^2\,.
\label{difdn1}
 \ea
Using Eq.~(\ref{ch2g}) to write $G(\phi_S)$ in the form 
\be
G(\phi_S) = \frac{\lm_1}{1 - \lm_2 \fp(\phi_L) (\chi_S / \chi_L)^{r p}}\,,
\label{gbca}
\ee
we get
\be
\frac{\partial G\(\phi_S\)}{\partial \phi_L} 
= p \lm_2 \phi_S^p \frac{\chi_S^2 G\(\phi_S\)}{\phi_L G\(\phi_L\)}\,.
\label{dgb2dpa}
\ee
Substituting the above relation into Eq.~(\ref{difdn1}) and performing an integration by parts similar to that for the last term in Eq.~(\ref{dnps}),
we obtain
\be
\frac{\partial \delta_{N_{LS}}}{\partial \phi_L}
= - 2 \frac{\delta_{N_{LS}} \lm_1}{\phi_L r G(\phi_L)}
+  G_{\phi_L} \chi_L^2 - p \lm_2 \phi_S^p \frac{\chi_S^2 G\(\phi_S\)}{\phi_L G\(\phi_L\)}\,.
\label{ddelnab2dpa}
\ee
From Eq.~(\ref{gbca}),
it can be shown that
\be
\frac{\partial G\(\phi_S\)}{\partial \chi_L} 
= - \frac{r \lm_2 p}{\chi_L \lm_1} \phi_S^p \chi_S^2 G\(\phi_S\)\,.
\label{dgb2dca}
\ee
Using the above relation and the same calculations as for Eq.~(\ref{ddelnab2dpa}),
we get
\be
\frac{\partial \delta_{N_{LS}}}{\partial \chi_L}
= 2 \frac{\delta_{N_{LS}}}{\chi_L}
+ \frac{r \lm_2 p}{\chi_L \lm_1} \phi_S^p \chi_S^2 G\(\phi_S\)\,.
\label{ddelnab2dca}
\ee
Performing similar calculations as above,
one can show that
\ba
\frac{\partial \delta_{N_{S}}}{\partial \phi_L}
&=& p \lm_2 \phi_S^p \frac{\chi_S^2 G\(\phi_S\)}{\phi_L G\(\phi_L\)}
- p \lm_2 \phi_u^p \frac{\chi_u^2 G\(\phi_u\)}{\phi_L G\(\phi_L\)}
- 2 \frac{\lm_1 \delta_{N_S}}{r \phi_L G\(\phi_L\)}\,,
\\
\frac{\partial \delta_{N_{S}}}{\partial \chi_L}
&=&
r \lm_2 p \phi_u^p \frac{\chi_u^2 G\(\phi_u\)}{\chi_L \lm_1}
- r \lm_2 p \phi_S^p \frac{\chi_S^2 G\(\phi_S\)}{\chi_L \lm_1}
+ 2 \frac{\delta_{N_S}}{\chi_L}\,.
\ea

\end{document}